# Absence of the Rashba effect in undoped asymmetric quantum wells

P.S. Eldridge [1], W.J.H Leyland [2], J.D. Mar [2], P.G. Lagoudakis [1], R. Winkler [3], O.Z. Karimov [1], M. Henini [4], D Taylor [4], R.T. Phillips [2] and R T Harley [1]

[1] School of Physics and Astronomy, University of Southampton, Southampton, SO17 IBJ, UK
[2] Cavendish Laboratory, Madingley Road, Cambridge CB3 OHE, UK
[3] Department of Physics, Northern Illinois University, DeKalb, IL 60115, USA
[4] School of Physics and Astronomy, University of Nottingham, Nottingham NG7 4RD, UK

*To an electron moving in free space an electric field appears as a magnetic field which interacts with and can reorient the electron's spin. In semiconductor quantum wells this spin-orbit interaction seems to offer the possibility of gate-voltage control in spintronic devices but, as the electrons are subject to both ion-core and macroscopic structural potentials, this over-simple picture has lead to intense debate. For example, an externally applied field acting on the envelope of the electron wavefunction determined by the macroscopic potential, underestimates the experimentally observed spin-orbit field by many orders of magnitude while Ehrenfest's theorem suggests that it should actually be zero. Here we challenge, both experimentally and theoretically, the widely held belief that any inversion asymmetry of the macroscopic potential, not only electric field, will produce a significant spin-orbit field for electrons. This conclusion has far-reaching consequences for the design of spintronic devices while illuminating important fundamental physics.*

Spin-orbit splittings induced by applied electric field in semiconductor heterostructures in principle enable a new generation of spintronic devices where electron spin is manipulated by external gate voltage. From a fundamental viewpoint this topic has been intensely debated [1] and continues to generate surprises. Using symmetry arguments, Bychkov and Rashba [2] predicted a spin-orbit splitting of the bands induced by any perturbation which breaks the inversion symmetry of the macroscopic potential (Structural Inversion Asymmetry, SIA) which includes applied electric field. However Ando and co-workers [3] pointed out that Ehrenfest's theorem requires that, for an electron in a stationary quantum state, the expectation value of total electric field, including the confining potential, must vanish, suggesting that spin-splitting induced by the electric field should also vanish. Experimentally, an electric field applied along the growth axis does induce a linear spin-splitting for conduction electrons in quantum wells [4] but it is found to be orders of magnitude greater than would be expected for an electron moving in the applied field alone[1]. These apparent contradictions illustrate the complications resulting from the additional presence of ion-cores and hetero-interfaces. The problem was analysed by Lassnig [5] who, taking these features into account, showed that Ehrenfest's theorem applies only to motion of the electron in the conduction band potential and that the spin-splitting of the electron states is caused by the potential of the valence band. At the present time it is widely assumed that any perturbation which breaks the macroscopic inversion symmetry in a quantum well will generate a significant Rashba-type spin-splitting[6, 7]. It is important to verify this assumption not only for fundamental reasons but because, if true, it offers the possibility to engineer built-in spin-orbit coupling in devices to supplement or replace that from applied gate voltage.

Here we show, through ultra-fast optical measurements of temporal and spatial spin dynamics, that electrons in asymmetrically grown quantum wells do not have measurable Rashba-type spin-splittings. We also point out how this surprising result can be deduced from Lassnig's theoretical analysis[1]. Thus we conclude that in general it is not possible to engineer the spin-splitting in undoped heterostructures by built-in asymmetry of the confinement potential; a real electric field (i.e. a Hartree potential gradient) is also required for production of a Rashba spin-splitting. This can come from voltage applied externally or asymmetric separation of free charges within the structure, for example as a result of delta-doping.

**Experimental determination of Rashba splitting**

The spin-orbit splitting for an electron with momentum **p** can be represented by a vector $\Omega(\mathbf{p})$ whose magnitude and direction represents the precession of the spin in the effective magnetic field seen by the moving electron. Our measurement technique exploits the fact that when relaxation of a spin-polarised population of electrons occurs via the so-called Dyakonov-Perel[8, 9] mechanism the spin-relaxation rate along an axis, $i$, is given by

$$\frac{1}{\tau_{s,i}} = \tau_p^* \langle \Omega_\perp^2 \rangle \qquad (1)$$

where $\tau_p^*$ is the momentum scattering time of an electron and $\langle \Omega_\perp^2 \rangle$ is the mean square component of $\Omega(\mathbf{p})$ perpendicular to the axis $i$ taken over the spin-polarised electron population. In our samples we measure, separately but under the same conditions, the spin-relaxation rate, $\tau_{s,z}^{-1}$, along the growth axis, $z$, by time-resolved Kerr rotation [10] and the scattering time, $\tau_p^*$, by a time-resolved spin-grating technique[11, 12]. The ratio of these quantities gives $\langle \Omega_{x,y}^2 \rangle$, the mean squared component of $\Omega(\mathbf{p})$ in the quantum well plane. Rashba spin-orbit coupling corresponds to an in-plane component of $\Omega(\mathbf{p})$ and thereby increases the spin-relaxation rate.

In zincblende structure semiconductors, in addition to any Rashba-type spin-splitting, $\Omega_{SIA}(\mathbf{p})$, which is always oriented in the quantum well plane (Fig. 1), there exists an intrinsic component of spin-orbit splitting, $\Omega_{BIA}(\mathbf{p})$, due to inversion asymmetry of the underlying crystal structure (Bulk Inversion Asymmetry, BIA)[1]. The total vector $\Omega(\mathbf{p})$ is the sum of components, $\Omega_{SIA}(\mathbf{p}) + \Omega_{BIA}(\mathbf{p})$. In the standard (100)-oriented quantum wells (Fig. 1a), $\Omega_{BIA}(\mathbf{p})$ also lies in the quantum well plane rendering the spin-relaxation along the growth axis relatively insensitive to the Rashba component. For (110)-oriented quantum wells (Fig. 1b), symmetry dictates that $\Omega_{BIA}(\mathbf{p})$ lies along the growth axis for all electrons and, in principle, the only contribution to the spin-relaxation comes from the Rashba (SIA) component[1,13]. For symmetrically grown quantum wells, where the mirror symmetry is preserved, $\Omega_{SIA}(\mathbf{p})$ should be zero and the Dyakonov-Perel mechanism should be totally suppressed giving greatly extended spin-relaxation times. The linear dependence of $\Omega_{SIA}(\mathbf{p})$ on electric field will induce a quadratic increase of spin-relaxation rate when electric field is applied and, furthermore, any additional spin-orbit splitting induced, for example, by built-in asymmetry should be detected as an increase of the spin-relaxation rate.

In our measurements we use (110)-oriented AlGaAs/GaAs/AlGaAs quantum wells which are undoped to avoid Hartree potential gradients due to dopants and free carriers. For symmetrical wells we find that the Dyakonov-Perel relaxation is indeed strongly suppressed, the residual spin-relaxation being clearly determined by a different mechanism[14, 15]. Application of an electric field of only 80 kV/cm produces quadratic enhancement of the spin-relaxation rate and easily measurable spin-splittings (up to 200 $\mu eV$). To our surprise, for asymmetric wells having one abrupt and one graded interface we also find strong suppression of the Dyakonov-Perel mechanism even though the potential gradient corresponds to electric field greater than 100 kV/cm. Our measurements give an upper limit, $\leq 40 \mu eV$, for the r.m.s Rashba splitting $(\langle \Omega_{SIA}^2 \rangle)^{0.5}$.

**Samples, techniques and results**

We present measurements made on three different multiple-quantum-well samples, each grown by molecular beam epitaxy on a semi-insulating (110)-oriented GaAs substrate. Samples A and B are completely undoped with asymmetric and symmetric alloy engineering respectively (see Fig. 2a and b). Sample A is made up of 5 repeats of 8nm GaAs quantum wells each with a 30nm graded upper interface where the aluminium fraction (x) is varied from 0.04 to 0.4 (Fig. 2a), followed by a 12nm barrier. Sample B comprises 20 repeats of 7.5nm GaAs quantum wells with abrupt 12nm $Al_{0.4}Ga_{0.6}As$ barriers (fig 2b). Sample C is a *p-i-n* device structure (Fig.2c) with undoped quantum wells nominally identical to those of sample B grown in the insulating (*i*) region. This allows uniform field to be applied to the quantum wells which can be calculated from the applied voltage and layer thicknesses of the *i* region. The top and bottom layers are $p^+$ and $n^+$ respectively each separated by a 100nm layer of undoped AlGaAs from the multi-quantum-well stack. Figure 2d shows the current-voltage (*I-V*) characteristics of the 400μm diameter mesa device. In the dark there is normal diode behaviour with rapid onset of conduction in forward bias, above ~1.3 V, and very low leakage current in reverse bias. Illuminated under the conditions of an experiment, ie ~400μW of light focussed to a 60μm spot and resonant with the *n*=1 interband transition, there is a small increase of reverse bias current up to -3V (~80kV/cm), corresponding to a loss of no more than ~2% of the photoexcited carriers from the wells.

Figure 3 shows the measured values of the momentum relaxation time $\tau_p^*$ at different temperatures for samples A, B and C obtained using the transient spin-grating technique [11,12]. The sample is excited by optical pulses of 200 fs duration from a mode-locked Ti-sapphire laser at resonance with the *n*=1 heavy-hole valence to conduction band transition of the quantum wells. Two simultaneous pump pulses having orthogonal polarisations are incident at ± 4.1 degrees to the normal to the sample and interfere to produce a periodic modulation of circular polarisation corresponding a grating of spin-polarisation with pitch $\Lambda \approx 5.7$ μm. With orthogonal polarisations, interference does not produce fringes of intensity so that the photoexcited population (~$10^9$ cm$^{-2}$) varies smoothly across the pumped spot. The temporal decay of this spin-grating is plotted out by measuring first-order backward diffraction of a weak, normal incidence, delayed probe pulse. On a picosecond timescale, the decay rate of the grating is [11, 12]

$$\Gamma = D_s \frac{4\pi^2}{\Lambda^2} + \frac{1}{\tau_{s,z}} + \frac{1}{\tau_r} \qquad (2)$$

and that of the diffracted intensity is $2\Gamma$ (inset Fig. 3). $D_s$ is the electron spin-self-diffusion coefficient and $\tau_r$ the recombination time; the photoexcited holes lose their spin-polarisation on a sub-picosecond timescale and do not affect the observed decay [11,16]. The spin-relaxation and recombination times are known from the time-resolved Kerr measurements described below allowing determination of $D_s$. For undoped samples and with low photoexcited electron concentration, we may ignore electron-electron scattering and therefore approximate the spin-self-diffusion to particle-self-diffusion. Thus the ensemble momentum relaxation time $\tau_p$ for the electrons can be obtained using the Einstein formula $\tau_p = (m^*/k_B T) D_s$ and the scattering time $\tau_p^*$ is equal to $\tau_p$ [17].

The solid line in Figure 3 indicates $\tau_p^* \sim 1/T$ and is a reasonable fit to the points from the different samples. It represents the expected variation for a non-degenerate two-dimensional electron system, i.e. with energy-independent density of states, and dominant phonon scattering[18]. In such an intrinsic scattering regime we expect all the samples to have essentially the same scattering time at a given temperature and, within the uncertainties in the measurements, this appears a reasonable assumption. Accordingly, we use values read from this line to combine with the measured spin-relaxation times to obtain the spin-splittings.

Figure 4 shows examples of time-resolved Kerr rotation measurements on the three different samples at 200K. In these measurements, a single pump beam of circularly polarised picosecond optical pulses at close to normal incidence excites a small population ($\sim 10^9$ cm$^{-2}$) of conduction electrons with spins aligned along the growth axis. A linearly-polarised beam of weaker, delayed probe pulses monitors the temporal spin and population dynamics of the electrons through measurements of polarisation rotation (Fig. 4a) and change of intensity in reflection[10]. The time evolution of the spin-polarisation in the symmetric (B) and the asymmetric (A) undoped quantum wells are very similar. The spin-relaxation times are both longer than 1 ns, suggesting that the graded interface has little impact on the spin dynamics and hence the spin-splitting. Sample C with -3V applied (80kV/cm) has spin lifetime of less than 100ps, an order of magnitude smaller than the lifetimes measured in the other two samples and similar to that observed in comparable (100)-grown quantum wells. Therefore it is already clear that the electric field produces an easily detectable spin-splitting whereas built-in asymmetry does not.

Figure 4b shows the spin-relaxation rate in sample C as a function of the square of applied electric field at 170 K; the same form of variation was observed at all temperatures from 80K to 230K[12]. Above ~30 kV/cm, the variation is accurately linear, consistent with dominance of the Dyakonov-Perel spin-relaxation mechanism with spin-splitting linear in field. Here the combination of the transient Kerr rotation and spin-grating measurements can give a direct determination of the Rashba spin-splitting. Below ~30 kV/cm the spin-relaxation rate becomes approximately constant. Since the rate extrapolates to the origin from high field, this low-field relaxation *cannot* be due to the Dyakonov-Perel mechanism with a field-independent Rashba spin-splitting, for example due to accidental built-in asymmetry associated with interface roughness as we suggested in an earlier paper [4]. Furthermore the dependences of the low-field spin-relaxation on both temperature and excitation density rule out the Dyakonov-Perel mechanism altogether and suggest that it is related to an alternative mechanism, possibly Elliott-Yafet or Bir-Aronov-Pikus [14, 15]. Thus it is apparent that, as the field is reduced, the Rashba spin-splitting becomes sufficiently small that Dyakonov-Perel is no longer the dominant spin-relaxation mechanism and our measurement technique can only set an upper limit on the true Rashba splitting.

Figure 5 shows the r.m.s. Rashba spin-splitting obtained for the different samples between 80K and 300K. In sample C at 80 kV/cm the splitting increases approximately linearly with temperature. The dotted curve is essentially a guide to the eye; for a Boltzmann distribution of electron energies we expect that the r.m.s. splitting will increase as $T^{0.5}$ but in an earlier publication we have interpreted the more rapid observed increase as due to a temperature dependence of the electric-field-splitting coefficient [12]. In samples A and B and in sample C with zero field applied, the measured spin-splitting is roughly constant over the temperature range. As stated above this insensitivity to temperature is not characteristic of the Dyakonov-Perel mechanism and the values must be taken as an upper limit for the true Rashba spin-splitting. The variation of the limit between samples can be traced to a dependence of the spin-relaxation on electron confinement energy, as was also observed by Ohno et al. [14]. The striking fact is that sample A which has asymmetric quantum wells, in which there is a potential gradient in the conduction band equivalent to greater than 100 kV/cm, has the lowest upper limit of the three samples, $(<\Omega_{SIA}^2>)^{0.5} \leq 40 \mu eV$.

**Theoretical considerations**

The following argument based on the envelope function approximation, indicates why for any sample there should be no measurable spin-splitting when the electric field is zero. Spin-splitting induced by inversion asymmetry reflects the coupled motion of electrons in the conduction band and holes in the valence band [1,5]. When these motions are decoupled (e.g., by means of a unitary transformation of the Hamiltonian) the Hamiltonian acting in the subspace of the electron states aquires an extra term, which is precisely the Rashba term

$$H_R = \alpha \, \partial V_v \, (\sigma_x p_y - \sigma_y p_x) \qquad (3)$$

Here $\sigma_x$ and $\sigma_y$ denote Pauli spin matrices, and $\partial V_v$ is the z-component of the gradient of the potential of the valence band $V_v$. The prefactor $\alpha$ is material-specific, but is independent of the geometry of an individual sample. We note that the Rashba term is fully analogous to the Pauli term that emerges in the context of the Dirac equation for relativistic electrons when the motion of the particles is decoupled from the motion of the antiparticles.

The important point to note about Eq. (3) is that, to obtain the conduction band spin-splitting, we take the expectation value, for the conduction band states, of the gradient of the potential in the *valence* band not, as might be expected, that in the conduction band, $\partial V_c$, which is not relevant for electron spin-splitting. Indeed according to Ehrenfest's theorem, for an electron in the conduction band the expectation value of $\partial V_c$ vanishes, i.e. in a stationary state the potential seen by the electron cannot exert a force.

In general, $V_v$ and $V_c$ contain both the Hartree-like contribution $V_H$ from the charge distribution in the system plus any externally applied voltage as well as contributions $V_{v,int}$ and $V_{c,int}$ from the position-dependent band edges in the sample. The gradient $\partial V_H$ is the same in both the conduction and valence bands while $\partial V_{v,int}$ is equal to $\partial V_{c,int}$ times a constant factor, $\Sigma$, the ratio of the valence and conduction band offsets. Note that for GaAs/AlGaAs $\Sigma \sim -0.54$ [19] and that the condition $\Sigma = +1$ corresponds to absence of hetero-interfaces in the structure.

Since the expectation value $<\partial V_c> = <\partial(V_H + V_{c,int})>$ vanishes by Ehrenfest's theorem, we have $<\partial V_H> = -<\partial V_{c,int}>$. This will be true for any sequence of interfaces, symmetric or asymmetric; the electron wavefunction will become more or less asymmetric to ensure this relationship. Consequently we can always write

$$<\partial V_v> = <\partial(V_H + V_{c,int})> = <\partial(V_H + \Sigma V_{c,int})> = (1-\Sigma)<\partial V_H> \qquad (4)$$

This first-order perturbation analysis indicates that the Rashba spin-splitting will always be directly proportional to $\partial V_H$ and so will be zero when there is no applied voltage and/or gradient of Hartree potential. It also contains the important point that the Rashba splitting vanishes if $\Sigma = 1$, ie if there are no hetero-interfaces in the structure.

To support this argument, we have made numerical calculations for three different GaAs/Al$_x$Ga$_{1-x}$As quantum well potentials based on diagonalisation of the 8x8 Kane Hamiltonian, including all higher-order effects and taking account of variations of all parameter values with alloy composition [1]. For a 10nm GaAs well with abrupt interfaces having one barrier with x = 1 and the other x = 0.2 we obtain a zero-field spin-splitting per unit wavevector 24μeV.nm and for the structure of sample A (see Fig. 2a) we obtain a spin-splitting 0.91μeV.nm. These values would give r.m.s spin-splittings at 300K of 5.4μeV and 0.2μeV respectively. These non-zero values are consistent with the Bychkov-Rashba symmetry argument [2] but are extremely small, far below the upper limit set by our measurements. The third calculation is for a structure with the same conduction band profile as A (Fig. 2a) but in which the valence band runs parallel to the conduction band in the 'graded' region. This corresponds to the right hand barrier in Figure 2a having a constant alloy composition, x = 0.04, plus a distribution of charge to give a Hartree potential with a gradient equivalent to that induced by alloy composition in A. The calculated splitting is 600μeV.nm corresponding to r.m.s splitting 135μeV at 300K. This would be readily measureable and reflects the presence of a real electric field. It is somewhat less than that measured in sample C at 80kV/cm corresponding to the fact that the electron probability density in the graded region is a relatively small fraction.

**Conclusions**

We conclude from our experimental measurements and theoretical argument that asymmetry of the alloy composition in a quantum well does not, by itself, generate a significant Rashba spin-orbit splitting for conduction electrons. There will only be a significant Rashba spin-splitting when there is a real electric field present. This may originate from external applied voltage or from an asymmetric distribution of charges in the structure. Our theoretical argument also shows that the presence of hetero-interfaces is a requirement for existence of a significant Rashba-type splitting. Thus we expect that *n-i-p-i* doping structures [20] in which the quantum wells are defined by in-built Hartree fields alone will show no significant Rashba splitting either for applied bias voltage or even if the doping profile lacks inversion symmetry. These results clarify a long-standing fundamental question concerning the spin-orbit coupling in heterostructures as well as helping to define the ground rules for the design of spintronic devices.

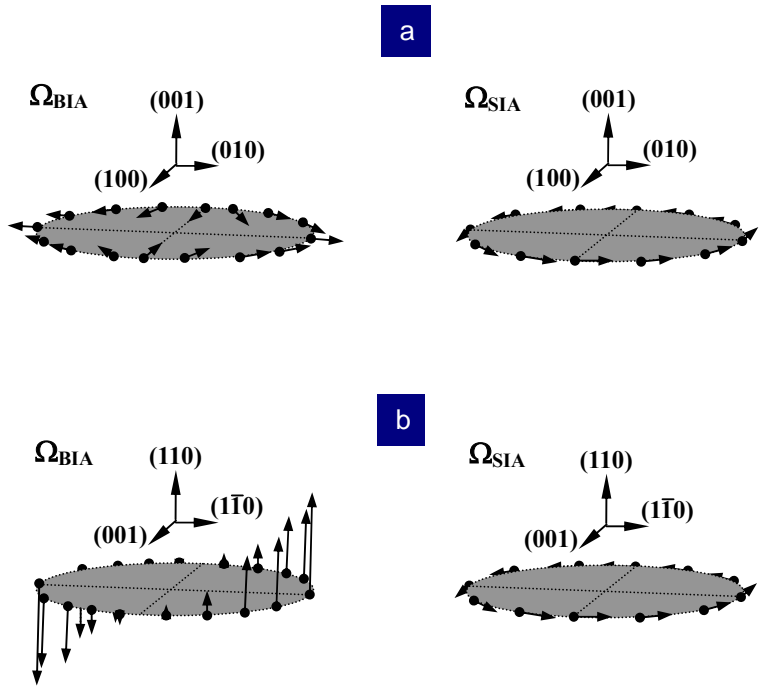

**Figure 1** Illustrating the magnitude and direction of BIA and SIA components of the spin-orbit splitting vector for electrons in (a) (001)- and (b) (110)-oriented quantum wells as a function of electron momentum. Momenta are confined in the quantum well plane and the vector is shown for momenta distributed on the perimeter of a disc; the growth axis ($z$) is vertical. To first approximation, the magnitude of the vector increases linearly with momentum.

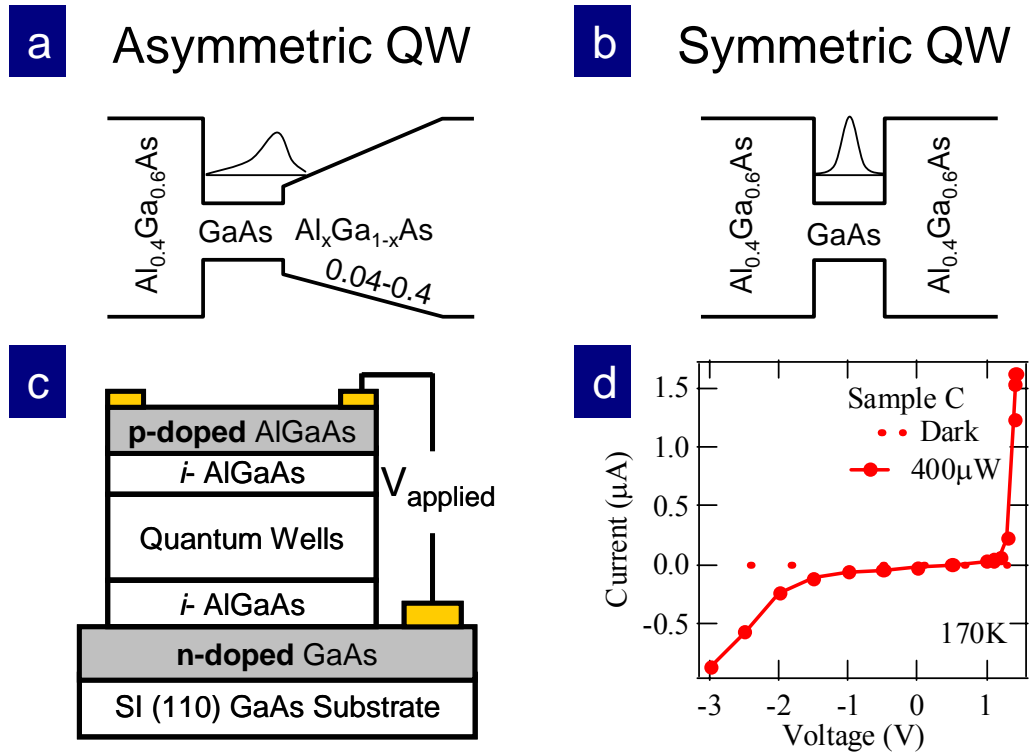

**Figure 2**  Band potential profiles and schematic representation of the electron probability density for undoped quantum wells in (a) sample A and (b) sample B . In A and B the GaAs wells are 8.0nm and 7.5nm thick respectively and in A the graded region is 30 nm thick. (c) Schematic of *p-i-n* device (sample C) used to apply variable electric field to nominally undoped quantum wells; symmetric quantum wells similar to those in sample B are incorporated in the *i*-region. (d) Current-voltage (*I-V*) characteristic of sample C in the dark and illuminated as for measurements of spin dynamics.

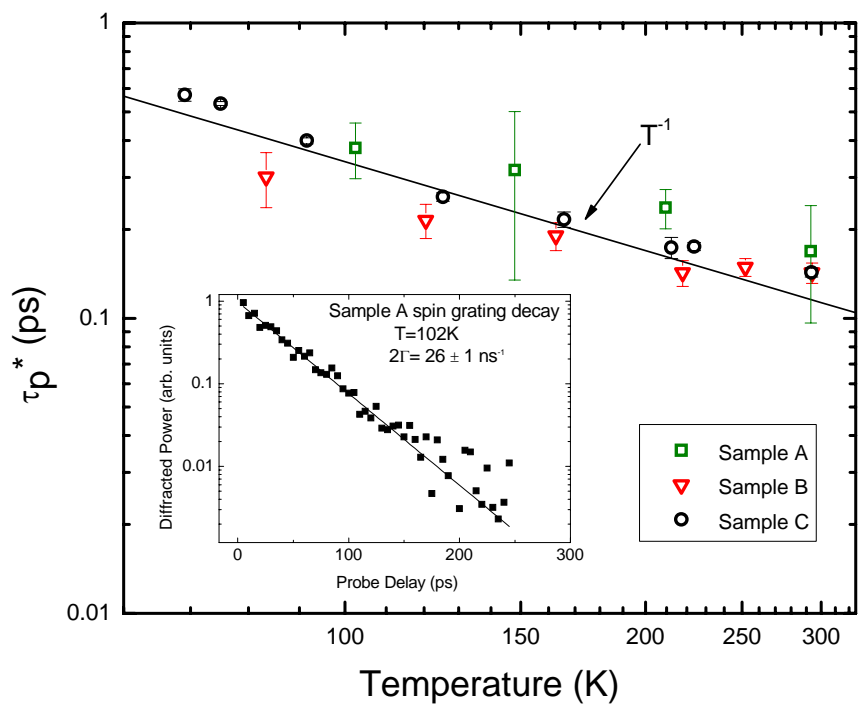

**Figure 3** Electron scattering time measured using spin-grating technique for samples A, B and C. The solid line is a $1/T$ fit to the points as discussed in the text. Inset is an example of a measured spin-grating decay.

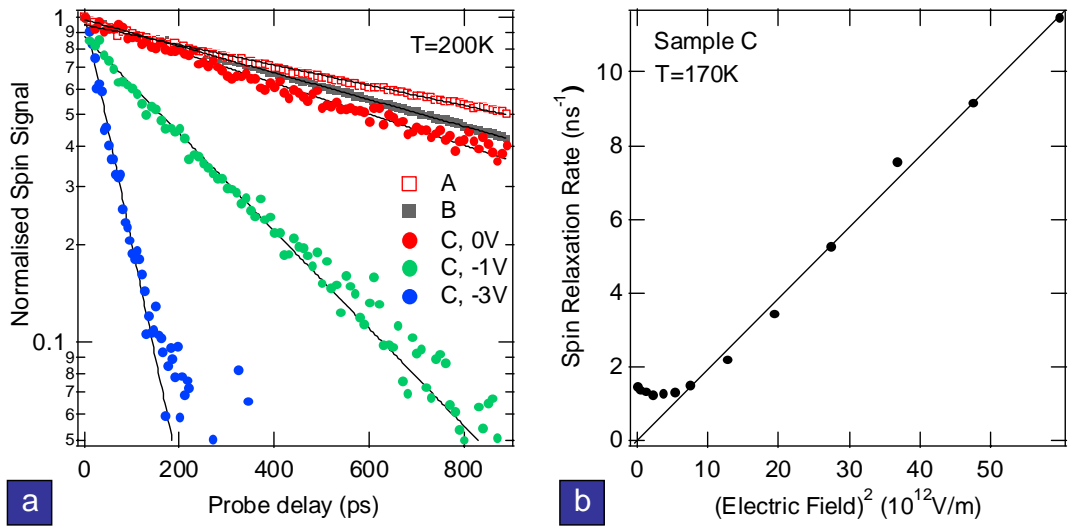

**Figure 4** (a) Decay of spin polarisation from time resolved Kerr rotation measurements at 200K in samples A (decay time 1397±12ps) and B (1058±10ps) and for sample C with three different bias voltages (0V, 914±13ps; -1V, 290±8ps; -3V, 62±5ps). (b) Variation of spin-relaxation rate with square of electric field in sample C at 170K. Note that above

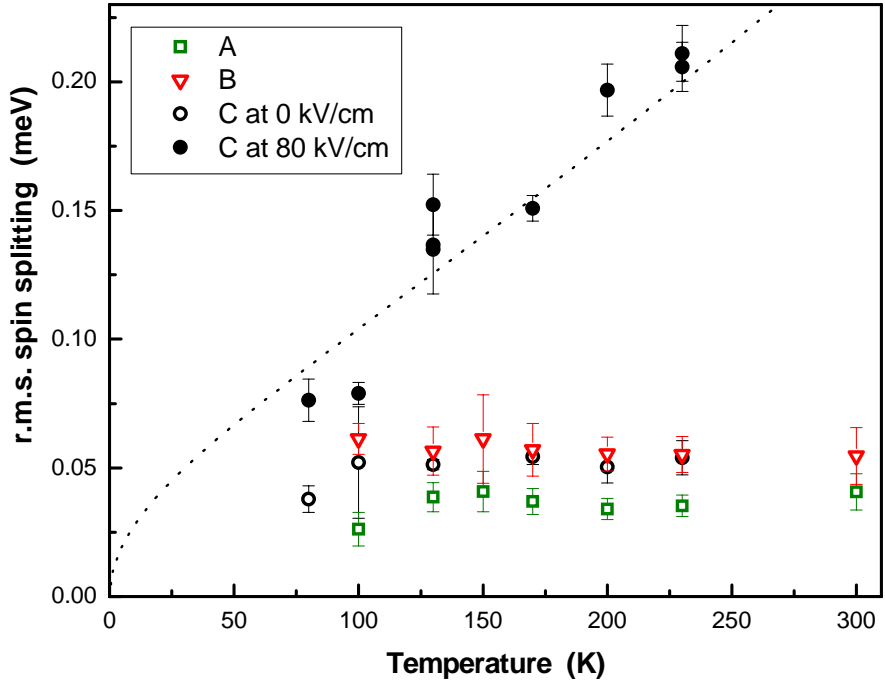

**Figure 5** Root mean square Rashba splitting for samples A, B and C obtained from time resolved Kerr and spin-grating measurements using eq. 1. For A, B and for C in zero electric field the measurements give an upper limit. For sample C at 80 kV/cm the points are a true measure of the splitting. Note that the r.m.s. Rashba splitting for sample A is less than 40 $\mu eV$ in spite of the fact that the built-in potential gradient in the conduction band corresponds to more than 100kV/cm.